\documentclass[fleqn]{article}
\usepackage{bm}
\usepackage{epsfig}
\usepackage{amsmath,amsfonts,amssymb}
\usepackage{color}
\usepackage{espcrc2}
\usepackage{graphicx}
\begin{document}
\title{Production of $\sigma$ in
$\psi(2S)\rightarrow \pi^{+}\pi^{-}J/\psi$ }
\author{\begin{small}
\begin{center}
M.~Ablikim$^{1}$,              J.~Z.~Bai$^{1}$,
Y.~Ban$^{12}$, X.~Cai$^{1}$,                  H.~F.~Chen$^{17}$,
H.~S.~Chen$^{1}$,              H.~X.~Chen$^{1}$,
J.~C.~Chen$^{1}$, Jin~Chen$^{1}$,                Y.~B.~Chen$^{1}$
% ...
\\
\vspace{0.2cm}
(BES Collaboration)\\
\vspace{0.2cm}
{\it
$^{1}$ Institute of High Energy Physics, Beijing 100049, People's Republic of China\\
% ...
$^{12}$ Peking University, Beijing 100871, People's Republic of China\\
% ...
$^{17}$ University of Science and Technology of China, Hefei 230026, People's Republic of China\\
% ...
}
\end{center}
\end{small}
}
\date{\today}
\begin{abstract}
  Using 14M $\psi(2S)$ events accumulated by BESII at
  the BEPC, a Covariant Helicity Amplitude Analysis is performed for
  $\psi(2S)\rightarrow\pi^+\pi^-J/\psi,~J/\psi\rightarrow \mu^+\mu^-$.
  The $\pi^+\pi^-$ mass spectrum, distinctly different from phase
  space, suggests $\sigma$ production in this process. Two different
  theoretical schemes are used in the global fit to the data. The
  results are consistent with the existence of the $\sigma$.
  The $\sigma$ pole position is determined to be
  $(552^{+~84}_{-106})-i(232^{+81}_{-72})$~MeV/$c^2$.
\end{abstract}
\maketitle
\normalsize
\setcounter{section}{0}
\setcounter{equation}{0}
\section{ \boldmath Introduction}
We report here a study of the process $\psi(2S) \to \pi^+ \pi^-
J/\psi$, which is the $\psi(2S)$ decay mode with the largest branching
fraction \cite{ref_42}, using very clean $\psi(2S) \to \pi^+ \pi^-
J/\psi~(J/\psi \to \mu^+ \mu^-)$ events.  Early investigations of this
decay by Mark I~\cite{marki} found that the $\pi^+ \pi^-$ mass
distribution is strongly peaked toward higher mass in contrast to what
is expected from phase space.  Furthermore, angular distributions
favored $S$-wave production between the $\pi\pi$ system and $J/\psi$,
as well as an $S$-wave decay of the dipion system.

BESI studied this process with much higher statistics (3.8 million
$\psi(2S)$ events) and found that an additional small $D$-wave
component was required in the decay of the dipion
system~\cite{rf_cut}.  Also various heavy quarkonium models
were fitted, and the parameters for these models
determined{~\cite{rf_cut}}.

Here, we fit the $\pi^+\pi^-$ system from $\psi(2S) \to \pi^+
\pi^- J/\psi$ decays with the $J^{PC}=0^{++}$ $\sigma$ meson.  In this
decay, the interaction between the $\pi\pi$ system and $\psi(2S)$ or
$J/\psi$ is small since these charmonium states are very narrow, so
the dipion system is a quasi-isolated system~\cite{ref_in}.

The $\sigma$ meson was introduced theoretically in the linear $\sigma$
model~\cite{ref_01}, and its existence was first suggested in a
one-boson-exchange potential model of nuclear forces~\cite{ref_11}.
The $\sigma$ meson is important due to its relation with dynamical
chiral symmetry breaking of QCD~\cite{ref_02}.

There was evidence for a low mass pole in early DM2~\cite{dm2}
and BESI~\cite{bes1-sigma} data on $J/\psi\rightarrow\omega\pi\pi$.
A huge event concentration in the $I = 0$ $S$-wave $\pi\pi$ channel
was observed in a $pp$ central production experiment in the region from
$m_{\pi\pi}=500$ to 600 MeV/$c^2$~\cite{pp}.  This peak is too
large to be explained as background~\cite{t-ishida}. Many studies on
the possible resonance structure in $\pi\pi$ scattering have appeared in
the literature~\cite{review}. It was proved that the existence of a
light and broad resonance is unavoidable even with non-linear
realization of chiral symmetry~\cite{xiao-1}. Careful theoretical
analyses were made to determine the pole location, which was found to
be $M-i\Gamma/2=(470\pm 30)-i(295\pm 20)$~MeV/$c^2$~\cite{roy-1} and
$M-i\Gamma/2=(470\pm 50)-i(285\pm 25)$~MeV/$c^2$~\cite{zhouzy}.
Renewed experimental interest arose from E791 data on
$D^+\rightarrow\pi^+\pi^-\pi^+$~\cite{e791}, where it was found that
$M=478^{+24}_{-23}\pm 17$~MeV/$c^2$, $\Gamma=324^{+42}_{-40}\pm
21$~MeV/$c^2$.  In the recent partial wave analysis of the decay
$J/\psi\rightarrow\omega\pi^+\pi^-$~\cite{ref_41} by BESII, the pole
position of $\sigma$ was determined to be $(541\pm 39)-i(252\pm
42)$~MeV/$c^2$.  All these experimental results still have large
uncertainties.

Fig.~\ref{ppj_int} shows the decay mechanism of
$\psi(2S)\rightarrow\pi^+\pi^-J/\psi$ in the $S$-matrix formalism.
There are three main contributions including an $S$-wave resonance
($\sigma$), a $D$-wave term~($2^+$), and a contact term which is the
destructive background required by chiral symmetry~\cite{ishida}.  The
total amplitude is the sum of these three components.  The decay
$\psi(2S)\rightarrow\pi^+\pi^-J/\psi$ can also be described with an
effective Lagrangian for the vector pseudo-scalar pseudo-scalar (VPP)
vertex, along with the $\pi\pi$ final state interaction (FSI) obtained
from $\pi\pi$ scattering data in a Chiral Unitary Approach
(ChUA)~\cite{guofk}.  In such an approach, the $\sigma$ resonance is
generated dynamically as a pole of the unitarized $t$-matrix, and the
pole position is $469-i203$~MeV$/c^2$~\cite{oset}.  A fit to the BESI
$\psi(2S)\rightarrow\pi^+\pi^-J/\psi$ data shows that the $\pi\pi$ FSI
plays an important role in this process~\cite{guofk}. {A similar
result was obtained in Ref.~\cite{guofk2} with a comparison of the
cases with and without the $\pi\pi$ FSI.  We fit our data with both
the $S$-matrix and ChUA schemes.

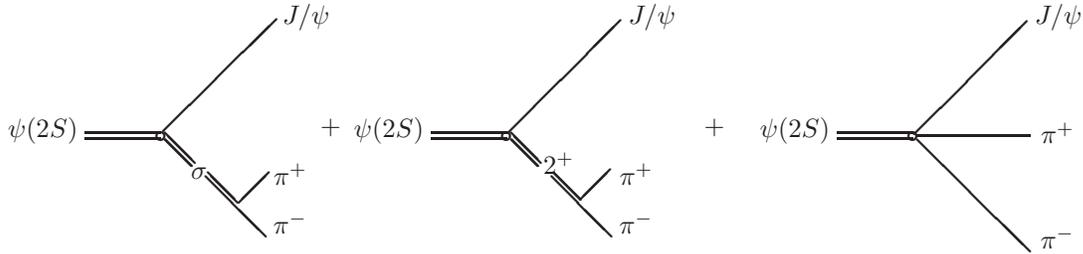
\begin{figure*}[htbp]
\centerline{
\setlength{\unitlength}{1.0cm}
\begin{picture}(14,5)\thicklines
\put(1.00,2.04){\line(1,0){1}} \put(1.00,1.96){\line(1,0){1}}
\put(2.00,2.00){\circle{0.10}} \put(2.04,2.06){\line(1,1){1.5}}
\put(2.06,2.04){\line(1,-1){0.40}}
\put(2.06,1.96){\line(1,-1){0.36}}
\put(2.63,1.51){\line(1,-1){0.40}}
\put(2.60,1.46){\line(1,-1){0.80}} 
\put(3.04,1.11){\line(1,1){0.40}}
\put(0.90,1.90){\makebox(0,0)[br]{$\psi(2S)$}}
\put(4.25,3.40){\makebox(0,0)[br]{$J/\psi$}}
\put(2.62,1.43){\makebox(0,0)[br]{$\sigma$}}
\put(3.95,1.35){\makebox(0,0)[br]{$\pi^+$}}
\put(3.95,0.70){\makebox(0,0)[br]{$\pi^-$}}
\put(4.40,2.0){\makebox(0,0)[br]{+}}
\put(5.60,2.04){\line(1,0){1}} \put(5.60,1.96){\line(1,0){1}}
\put(6.60,2.00){\circle{0.10}} \put(6.62,2.04){\line(1,1){1.5}}
\put(6.66,2.04){\line(1,-1){0.40}}
\put(6.66,1.96){\line(1,-1){0.36}}
\put(7.23,1.51){\line(1,-1){0.36}}
\put(7.20,1.46){\line(1,-1){0.80}} 
\put(7.58,1.15){\line(1,1){0.40}}
\put(5.50,1.90){\makebox(0,0)[br]{$\psi(2S)$}}
\put(8.85,3.40){\makebox(0,0)[br]{$J/\psi$}}
\put(7.50,1.50){\makebox(0,0)[br]{$2^+$}}
\put(8.55,1.35){\makebox(0,0)[br]{$\pi^+$}}
\put(8.55,0.70){\makebox(0,0)[br]{$\pi^-$}}
\put(9.50,2.0){\makebox(0,0)[br]{+}}
\put(11.00,2.04){\line(1,0){1}} \put(11.00,1.96){\line(1,0){1}}
\put(12.00,2.00){\circle{0.10}} \put(12.02,2.02){\line(1,1) {1.5}}
\put(12.06,2.00){\line(1,0) {1.5}}
\put(12.06,1.96){\line(1,-1){1.5}} 
\put(10.90,1.90){\makebox(0,0)[br]{$\psi(2S)$}}
\put(14.25,3.40){\makebox(0,0)[br]{$J/\psi$}}
\put(14.15,1.90){\makebox(0,0)[br]{$\pi^+$}}
\put(14.15,0.50){\makebox(0,0)[br]{$\pi^-$}}
\end{picture}}
\vspace{-0.8cm} \caption{Decay mechanisms for 
$\psi(2S)\rightarrow\pi^+\pi^-J/\psi$ in the $S$-matrix formalism. 
The final amplitude is the superposition of
$\psi(2S)\rightarrow\sigma J/\psi$, 
$\psi(2S)\rightarrow 2^+ J/\psi$, and 
$\psi(2S)\rightarrow(\pi^+\pi^-)_{\mbox{cont}}J/\psi$ $S$-matrix elements.}
\label{ppj_int}
\end{figure*}

\section{ \boldmath BESII Experiment}
The data sample used for this analysis is taken with the BESII
detector at the BEPC storage ring operating at the $\psi(2S)$
resonance. The number of $\psi(2S)$ events is $14.0\pm 0.6$
million~\cite{tot-num}, determined from the number of inclusive
hadrons.

The Beijing Spectrometer~(BES) detector is a conventional
solenoidal magnet detector that is described in detail in
Ref.~\cite{bes1}; BESII is the upgraded version of the BES
detector~\cite{bes2}. A 12-layer vertex chamber~(VC) surrounding
the beam pipe provides track and trigger information. A 40-layer main drift
chamber~(MDC), located radially outside the VC, provides trajectory
and energy loss~($dE/dx$) information for charged tracks over $85\%$
of the total solid angle. The momentum resolution is
$\sigma_{p}/p=0.017\sqrt{1+p^2}$~($p$ in GeV/$c$), and the $dE/dx$
resolution is $\sim 8\%$. An array of 48 scintillation counters
surrounding the MDC measures the time-of-flight~(TOF) of charged
tracks with a resolution of $\sim$ 200 ps for hadrons. Radially
outside the TOF system is a 12 radiation length, lead-gas barrel
shower counter (BSC). This measures the energies of electrons and
photons over $\sim 80\%$ of the total solid angle with an energy
resolution of $\sigma_{E}/E=22\%/\sqrt{E}$~($E$ in GeV). Outside the
solenoidal coil, which provides a 0.4 Tesla magnetic field over
the tracking volume, is an iron flux return that is instrumented
with three double layers of counters~(MUID) that identify muons of
momentum greater than 0.5 GeV/$c$.

A GEANT3 based Monte Carlo~(MC) simulation program~\cite{simbes} with
detailed consideration of detector performance (such as dead
electronic channels) is used to simulate the BESII detector. The
consistency between data and MC simulation has been carefully
checked in many high purity physics channels, and the agreement
is quite reasonable~\cite{simbes}.

\section{\boldmath Event Selection}
Events with $\pi^+\pi^-\mu^+\mu^-$ final states and with the
invariant mass $m_{\mu^+\mu^-}$ constrained to the $J/\psi$ mass are
selected for analysis. Each track, reconstructed using hits in the
MDC, must have a good helix fit in order to ensure a correct error
matrix in the kinematic fit, and the number of tracks are required to
be between 4 and 7.


To select a pair of muons, the muon pair candidates tracks are
required to have net charge zero; $p_{\mu^+}>1.3$~GeV/$c$ or
$p_{\mu^-}>1.3$~GeV/$c$ or $p_{\mu^+}+p_{\mu^-}>2.4$~GeV/$c$; $\vert
\cos\theta_\mu\vert <0.6$ to ensure that tracks are in the sensitive
region of the MUID; the cosine of the angle between these two tracks
in their rest frame $\cos\theta^{cm}_{\mu^+\mu^-}<-0.975$ to guarantee
the collinearity of the tracks; the sum of the MUID hits
$N^{hit}_{+}+N^{hit}_{-}\geq 3$ to ensure that the tracks are muons; and the
invariant mass of two candidate tracks $m_{\mu^+\mu^-}$ within
0.35~GeV/c$^2$ of the $J/\psi$ mass.

For $\pi^+\pi^-$ pair selection, the two candidate tracks are also
required to have net charge zero. Each track is required to have
momentum $p_{\pi}<0.5$~GeV/$c$, polar angle $\vert \cos\theta\vert
<0.75$, and transverse momentum $p_{\pi xy}>0.1$~GeV/$c$ to reject
tracks that spiral in the MDC.  The $dE/dx$ measurement of each track
must be within three standard deviations of the $dE/dx$ expected for
the pion hypothesis, and the cosine of the laboratory angle between
the candidate tracks must satisfy $\cos\theta_{\pi\pi}<0.9$ to
eliminate $e^+e^-$ pairs from $\gamma$ conversions. The mass recoiling
against the candidate $\pi^+\pi^-$ pair, $m^{recoil}_{\pi^+\pi^-}$, is
shown in Fig.~\ref{ppm-pprec}a.  In order to get well reconstructed
signal events and to suppress background, $\vert
m^{recoil}_{\pi^+\pi^-}-m_{J/\psi}\vert<20$~MeV/$c^2$, corresponding
to three times the mass resolution, is required.


With the above selection criteria, about 40,000
$\psi(2S)\rightarrow\pi^+\pi^-J/\psi\rightarrow\pi^+\pi^-\mu^+\mu^-$
candidate events are obtained. Fig. {~\ref{ppm-pprec}b} shows the
$\pi^+\pi^-$ invariant mass distribution for these events, where the
dots with error bars are data, and the histogram is Monte Carlo
simulation with the PPGEN generator, which is based on chiral symmetry
arguments and partially conserved axial vector currents~\cite{cahn}.
It describes the low mass $\pi\pi$ spectrum reasonably well but not
the high mass region; the inconsistency between data and Monte Carlo
will be considered in the systematic errors.


\begin{figure*}[htbp]
\centerline{\hbox{\psfig{file=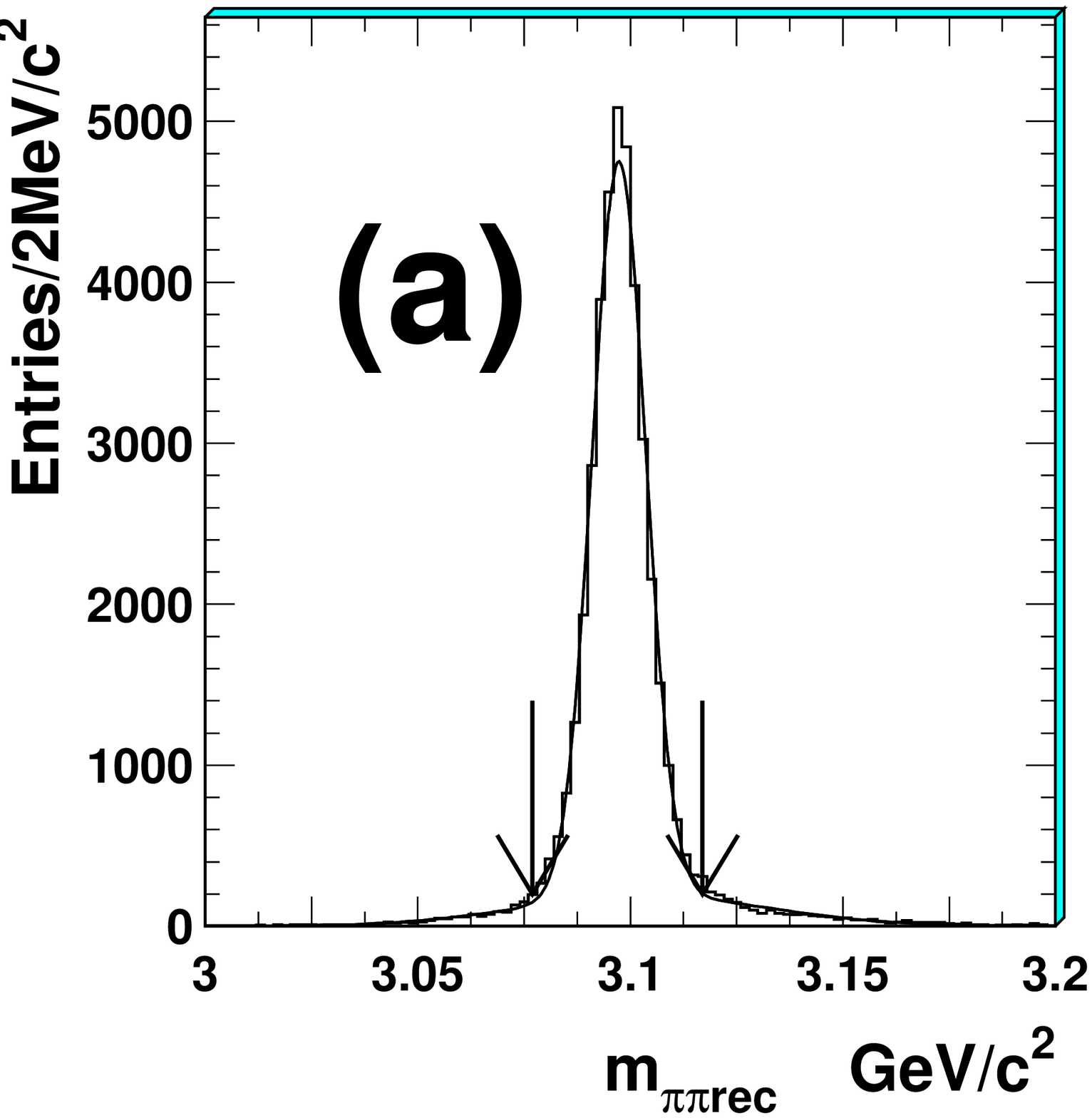,width=7cm,height=6.0cm}}
\hbox{\psfig{file=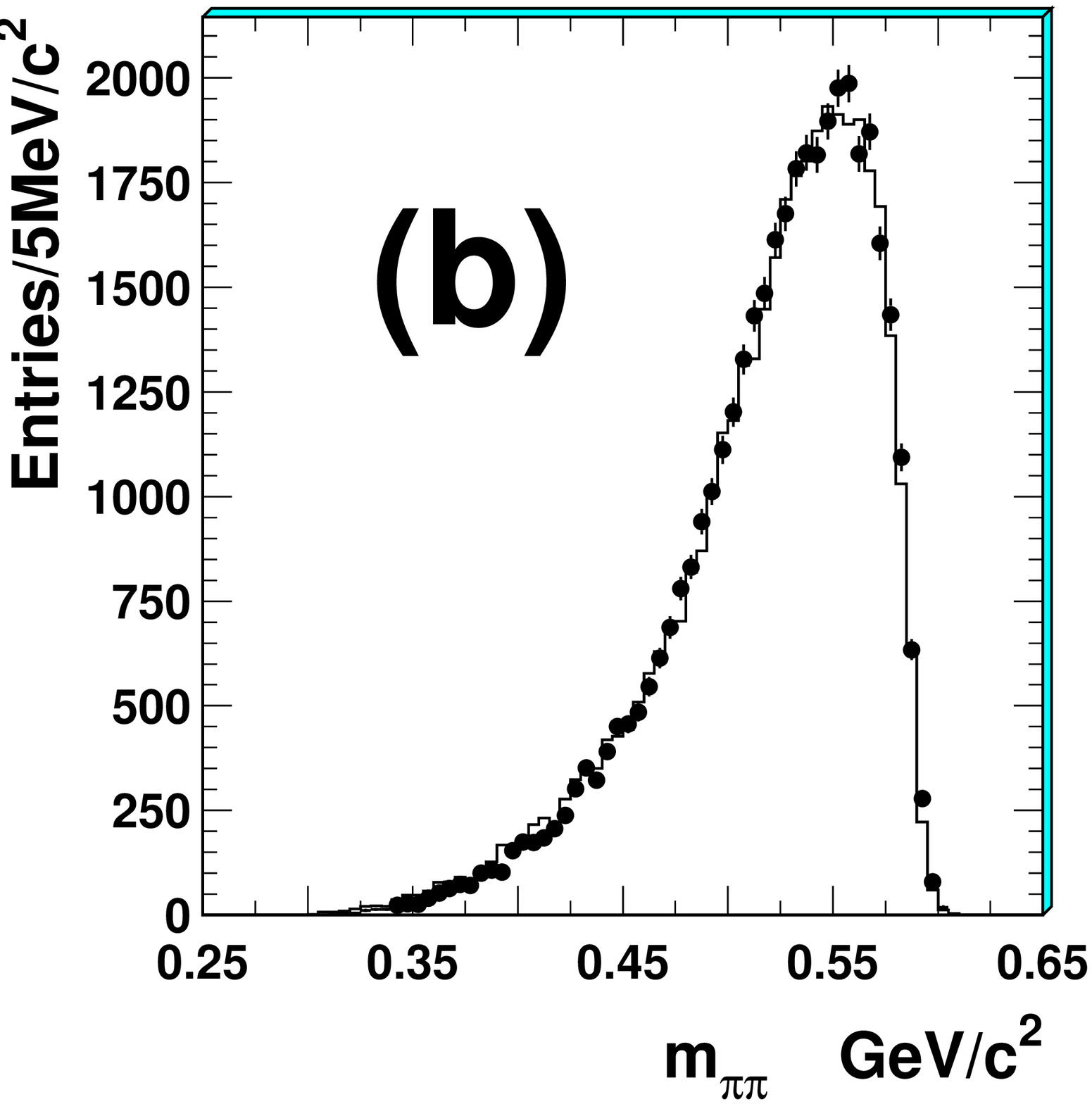,width=7cm,height=6.0cm}}}
\caption{
Distributions of candidate
$\psi(2S)\rightarrow\pi^+\pi^-J/\psi~(J/\psi\rightarrow\mu^+\mu^-)$
events. (a) is the $\pi^+\pi^-$ recoil mass spectrum,
fitted with a double Gaussian function, and
(b) is the $\pi^+\pi^-$ invariant mass spectrum.
The histogram in (a) is data, the curve is the fit,
the events between arrows are selected; in (b)
dots with error bars are data, and the histogram is MC simulation.}
\label{ppm-pprec}
\end{figure*}


The main background channels are $\psi(2S)\rightarrow\eta J/\psi
~(J/\psi\rightarrow\mu^+\mu^-)$,
$\psi(2S)\rightarrow\pi^+\pi^-J/\psi~(J/\psi\rightarrow\pi^+\pi^-)$
and $\psi(2S)\rightarrow\pi^+\pi^-J/\psi~(J/\psi\rightarrow\rho\pi)$.
However, MC simulation indicates that their total contribution is less
than 0.1\%, which can be neglected.  The contamination from continuum
production $e^+e^-\rightarrow\pi^+\pi^-\mu^+\mu^-$ is also very small
and neglected in this analysis.


\section{\boldmath Analysis Method}
Two different schemes are used to fit our data. In the first, based
on the diagrams in Fig.~\ref{ppj_int} and taking the VPP vertex
as a constant, the total differential cross section
which describes $\psi(2S)\rightarrow\pi^+\pi^-J/\psi$ is
\begin{eqnarray}
\frac{d\sigma}{d\Omega}&=&\sum_{M\lambda_{\psi}}\vert A\vert^2
\nonumber\\
&=&\sum_{M\lambda_{\psi}}
               \left|
                     A_{0}
                     +\sum_{\lambda_{2}}A_{2}
                     +A_{\mbox{contact}}
                    \right|^{2},
\end{eqnarray}
where $A_s$ represents the amplitude for
$\psi(2S)\rightarrow XJ/\psi\rightarrow\pi^+\pi^-J/\psi$ with
the spin of $X$ being $s$, $\Omega$ represents
the solid angle, $M$ is the magnetic quantum number
along $Z$-axis of $\psi(2S)$,
$\lambda_{\psi}$ and $\lambda_{2}$ are the helicities
of the $J/\psi$ and $2^+$ components, respectively, and
$A_{contact}$ is the amplitude of the contact term.

In the second, considering the VPP vertex and the $S$-wave $\pi\pi$
FSI, while neglecting the $D$-wave FSI, the amplitude
is~\cite{guofk}
\begin{eqnarray}
A = V_0+V_{0S}\cdot G\cdot 2t^{I=0}_{\pi\pi\rightarrow\pi\pi}~,
\\
{V_0 = -\frac{4}{f_\pi^2} (g_1p_1\cdot p_2+g_2p_1^0p_2^0+g_3m^2_\pi)
\epsilon^{*}\cdot\epsilon^{'}},
\end{eqnarray}

where $G$ is the two-pion loop propagator, {$V_{0S}$ is the S-wave
part of $V_0$,} and $t^{I=0}_{\pi\pi\rightarrow\pi\pi}$ is the
full S-wave {$I=0$ $\pi\pi\rightarrow\pi\pi$} $t$-matrix, which is
the same as those defined in Refs.~\cite{guofk,oset}; $p_1$
and $p_2$ are the four momenta of the two pions, and $p^0_1$ and
$p^0_2$ are their energies in the lab frame; $g_1$,
$g_2$, and $g_3$ are free parameters to be determined
 by data.

The normalized probability density function used to describe the
whole decay process is
\begin{eqnarray}
f(x,\alpha)=\frac{d\sigma/d\Omega}{\sigma},
\end{eqnarray}

where $x$ represents a set of quantities which are measured by
experiment, and $\alpha$ represents unknown parameters to be
determined. The total cross section, $\sigma$, can be expressed as
\begin{eqnarray}
\sigma=\int{\epsilon(\Omega)\frac{d\sigma}{d\Omega}d\Omega},
\end{eqnarray}
where $\epsilon(\Omega)$ is the detection efficiency which is
usually a function of detector performance. The total cross
section can be determined by MC integration. Re-weighting
a total of $N$ generated events based on simulated
$\psi(2S)\rightarrow\pi^+\pi^-J/\psi$ using a phase space generator,
the total cross section is then
\begin{eqnarray}
\sigma=\frac{1}{N_{mc}}\sum^{N_{mc}}_{i=1}
\left\{\frac{d\sigma}{d\Omega}\right\}_i.
\end{eqnarray}

where $N_{mc}(<N)$ is the number of MC simulated events
after applying the selection criteria.

The maximum likelihood function~\cite{ref_05,ref_07} is given by the
joint probability density of the selected $\psi(2S)\rightarrow
\pi^+\pi^-\mu^+\mu^-$ events,
\begin{eqnarray}
\hspace{2cm} {\cal L}=\prod_{i=1}^{N_{evt}}f(x,\alpha)~,
\end{eqnarray}
and a set of values, $\alpha$, is obtained by minimizing the
function $S$,
\begin{eqnarray}
\hspace{2cm} S=-\log{\cal L}~.
\end{eqnarray}

For the amplitudes in the first model,
the amplitudes for the cascade two-body decay process can be expanded with
helicity amplitudes as:
\begin{eqnarray}
A_{s}=F^{J}_{\lambda\nu}D^{*J}_{M,\lambda-\nu}
BW_{X}(S_{\pi\pi},m_X,\Gamma_{X})F^{s}_{00}D^{*s}_{\lambda,0}~,
\end{eqnarray}

where $F^{J}_{\lambda\nu}$ is the helicity amplitude,
which can be found in Ref.{~\cite{ref_07}},
$D^{J}_{M,\lambda-\nu}(\phi,\theta,0)$ is the
$D$-function, and $BW_{X}(S_{\pi\pi},m_X,\Gamma_{X})$ is the
Breit-Wigner propagator of $X$, defined as :
\begin{eqnarray}
BW_{X}(s_{\pi\pi},m_X,\Gamma_X)\hspace{3cm}\nonumber\\
\hspace{2cm}=\frac{1}{s_{\pi\pi}-m_X^2+im_X\Gamma_X(s_{\pi\pi})}~. \label{bw0}
\end{eqnarray}

The $\sigma$ particle, a broad structure
in the low $\pi^+\pi^-$ mass region,
is not a typical Breit-Wigner resonance. In the first model,
four types of Breit-Wigner parameterizations are used to describe
it:

\begin{itemize}
\item[] 1) Constant width
\begin{eqnarray}
 \Gamma_X(s)= \Gamma~.
\label{constant}
\end{eqnarray}

\item[]2) Width containing a kinematic factor, which was used by the
E791 Collaboration{~\cite{e791}}
\begin{eqnarray}
 \Gamma_X(s)=\rho\Gamma
=\sqrt{1-\frac{4m_{\pi}^2}{s}}\Gamma~.
 \label{rho}
\end{eqnarray}

\item[]3) P.K.U. ansatz{~\cite{zheng-hq-1}}, which removes the spurious
singularity hidden in Eq. (\ref{rho})
\begin{eqnarray}
 \Gamma_X(s)=\rho \frac{s}{m_X^2}\Gamma
=\sqrt{1-\frac{4m_{\pi}^2}{s}}\frac{s}{m_X^2}\Gamma~. \label{pku}
\end{eqnarray}

\item[]4) Zou and Bugg's approach{~\cite{zou}}, where the form includes
explicitly into $\Gamma_X(s)$ the Adler zero at $s = m^2_\pi/2$.
\begin {eqnarray}
\Gamma_X(s) &=& g_1\frac {\rho _{\pi \pi }(s)}{\rho _{\pi \pi
}(m_X^2)} + g_2\frac {\rho _{4\pi } (s)} {\rho _{4\pi }(m_X^2)} \\
g_1 &=& f(s)\frac {s - m^2_\pi /2}{m_X^2-m^2_\pi/2}e^{-\frac{s-m_X^2}{a}},
\nonumber
\end {eqnarray}
where the definitions of  $\rho _{\pi \pi }$, $\rho _{4\pi}$, and $f(s)$
are the same as Ref.~\cite{ref_41}.
\end{itemize}

\section{\boldmath Partial wave analysis}
The minimization used in the partial wave analysis and to obtain the
pole parameters of the $\sigma$ is based on MINUIT~\cite{minuit}.  For
the first model, the components considered include amplitudes of
$\sigma(0^+)$, a $D$-wave term, and a contact term. The tail of the
$f_{0}(980)$ has been tried in the fit. However, it has similar
behavior to the contact term in this mass region, and therefore it is
ascribed to the contact term.  All four $\sigma$ Breit-Wigner
parameterizations fit the data well, but have strong destructive
interference with the contact term, especially in the low
$\pi^{+}\pi^{-}$ invariant mass region.  The $D$-wave contribution is
only 0.3 to 1\%, in agreement with the BESI result~\cite{rf_cut} based
on a different analysis method. Fig.{~\ref{mass-just}} shows the
projections of the fit results compared with data for the Breit-Wigner
parameterization for the P.K.U. ansatz; other parameterizations give
similar results.

\begin{figure*}[htbp]
\centerline{
\psfig{file=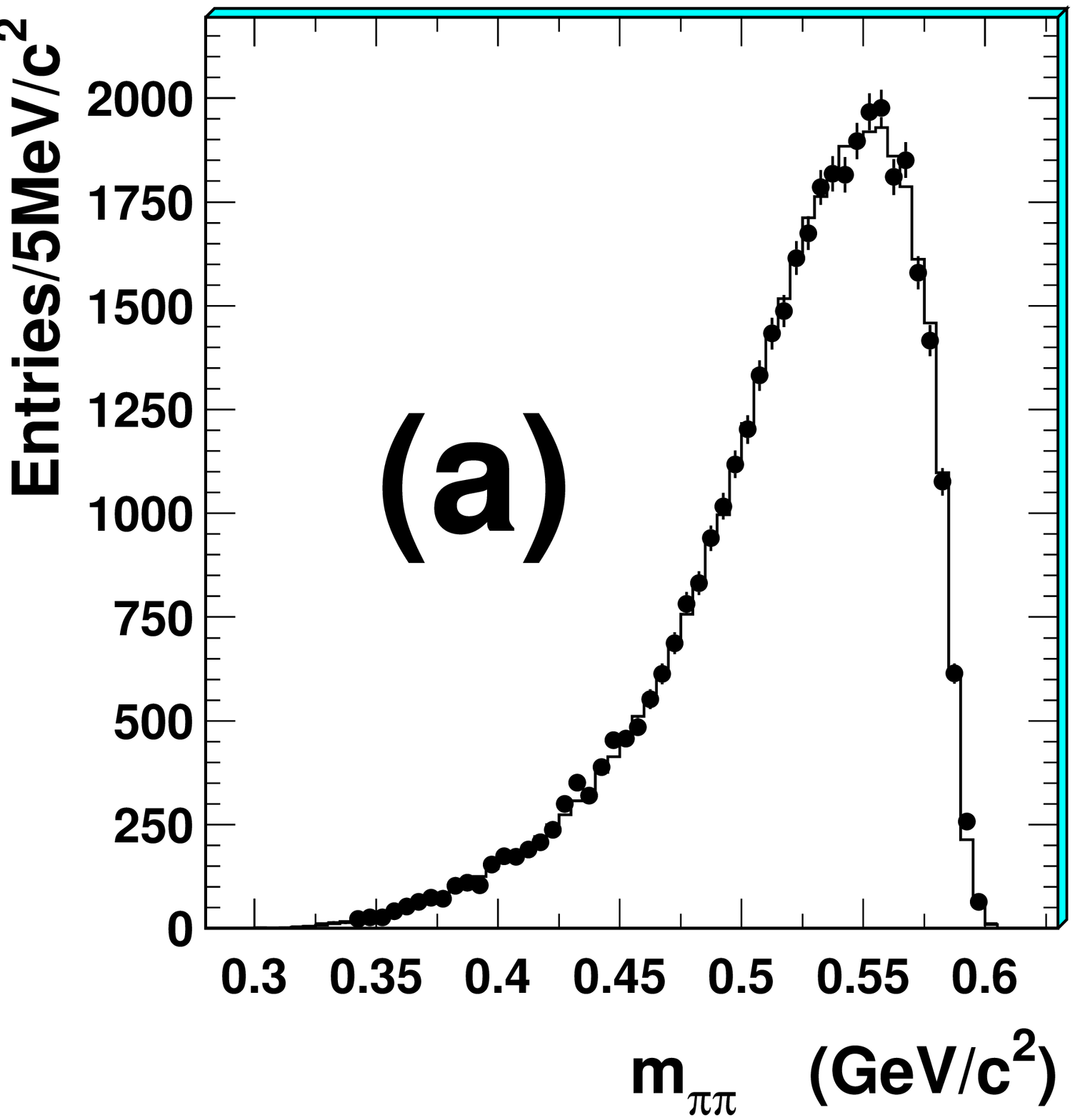,   width=4.5cm,height=4.0cm}
\psfig{file=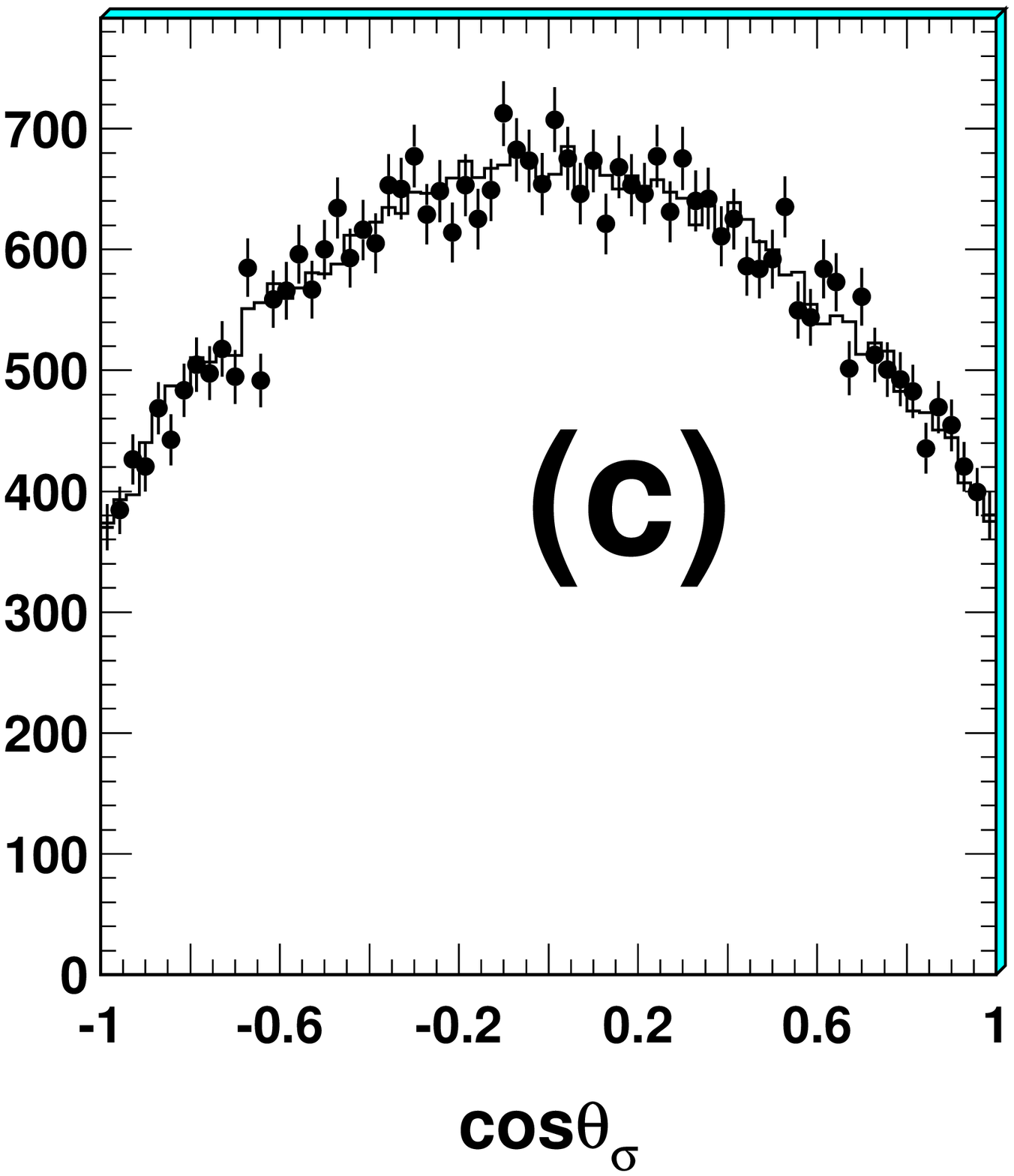,  width=4.5cm,height=4.0cm}
\psfig{file=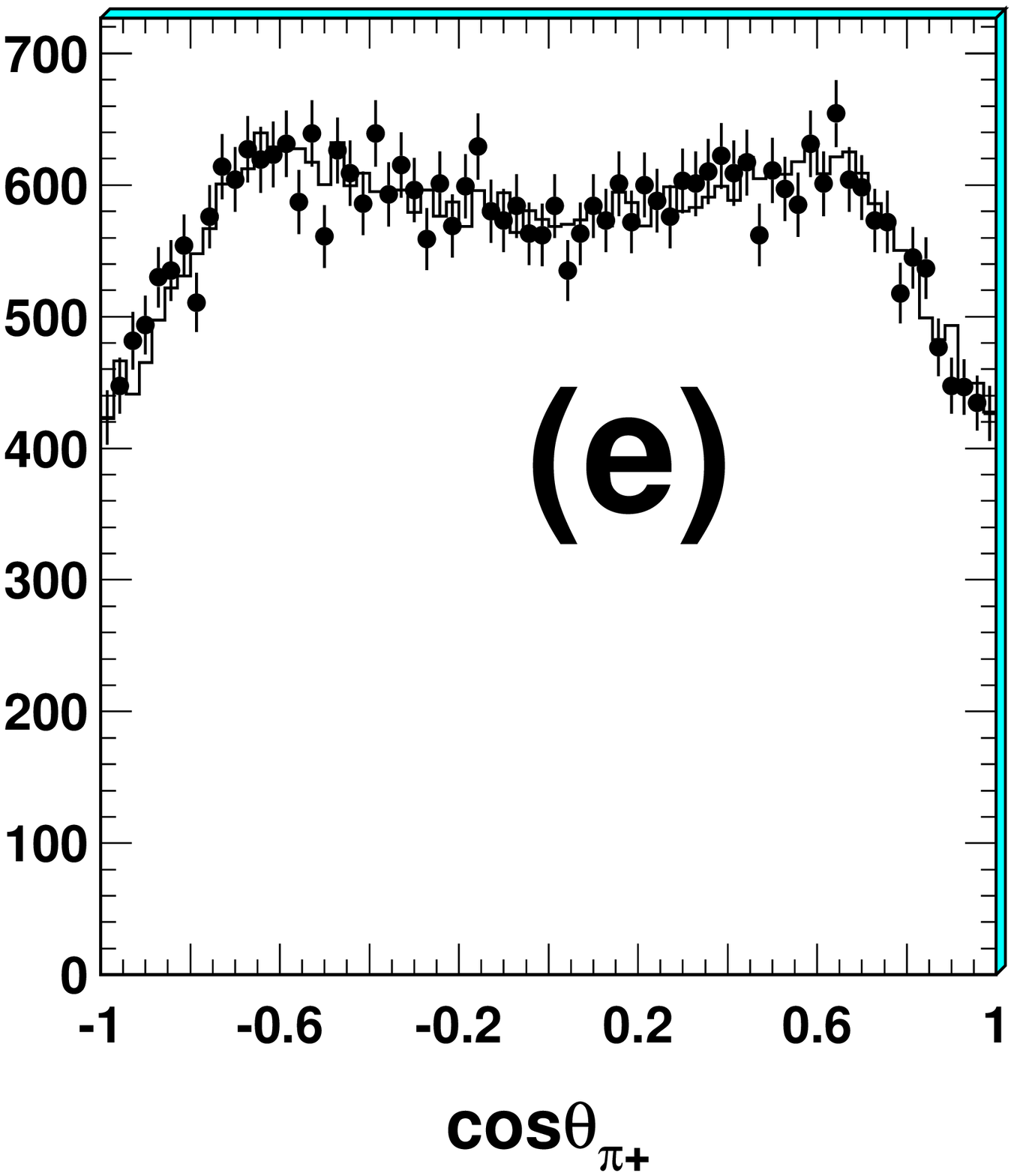,  width=4.5cm,height=4.0cm}}
\centerline{
\psfig{file=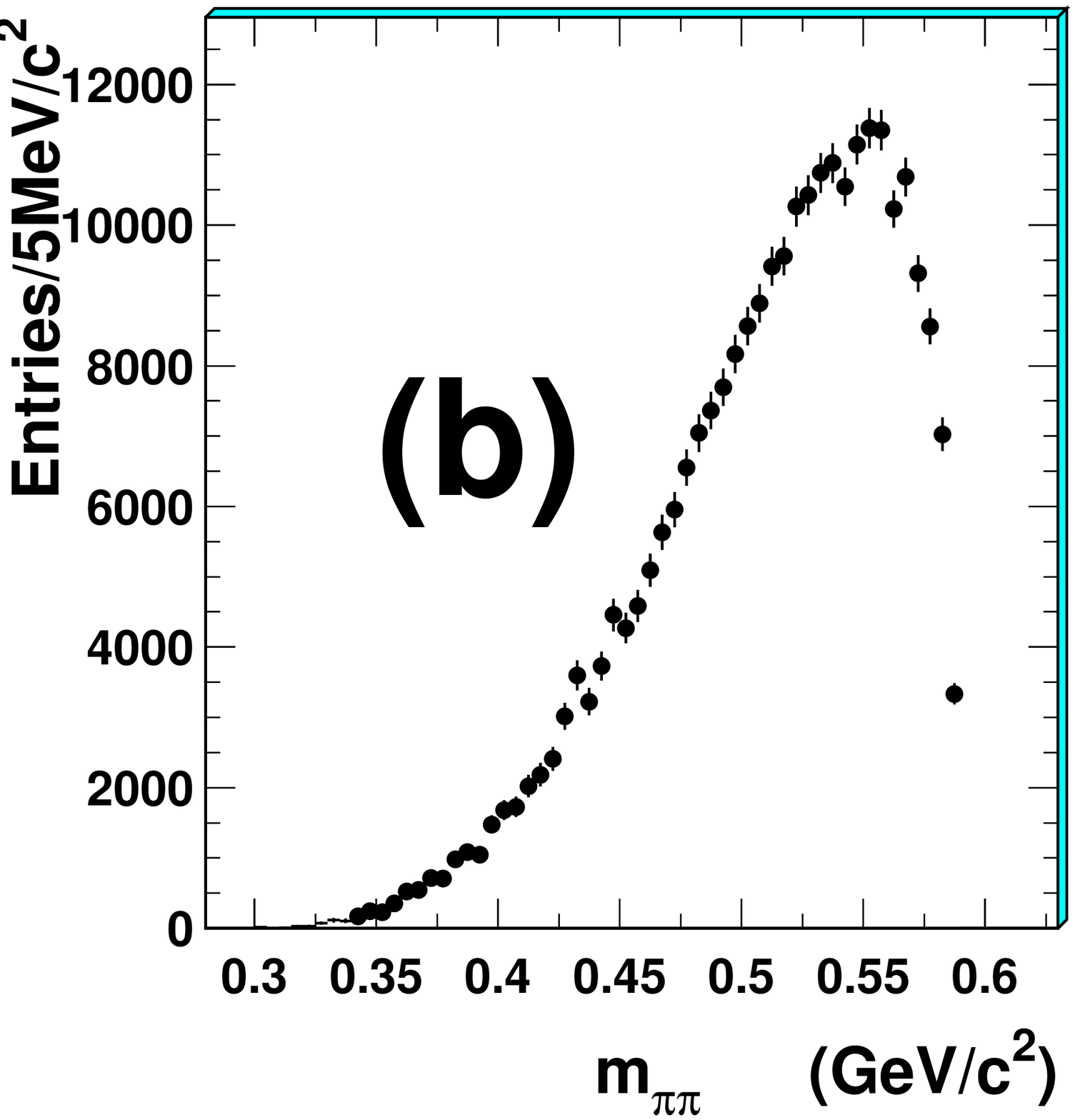,  width=4.5cm,height=4.0cm}
\psfig{file=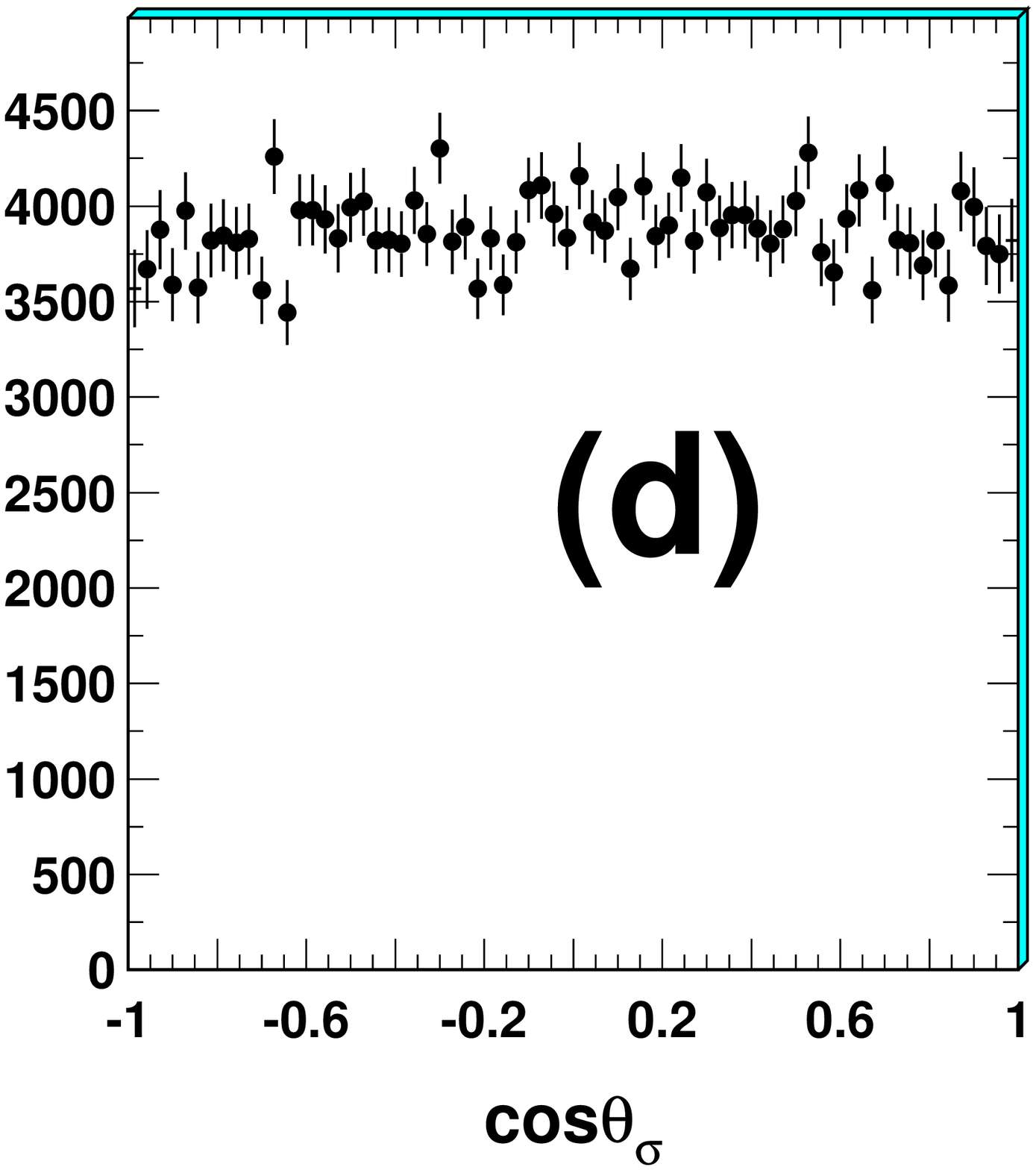, width=4.5cm,height=4.0cm}
\psfig{file=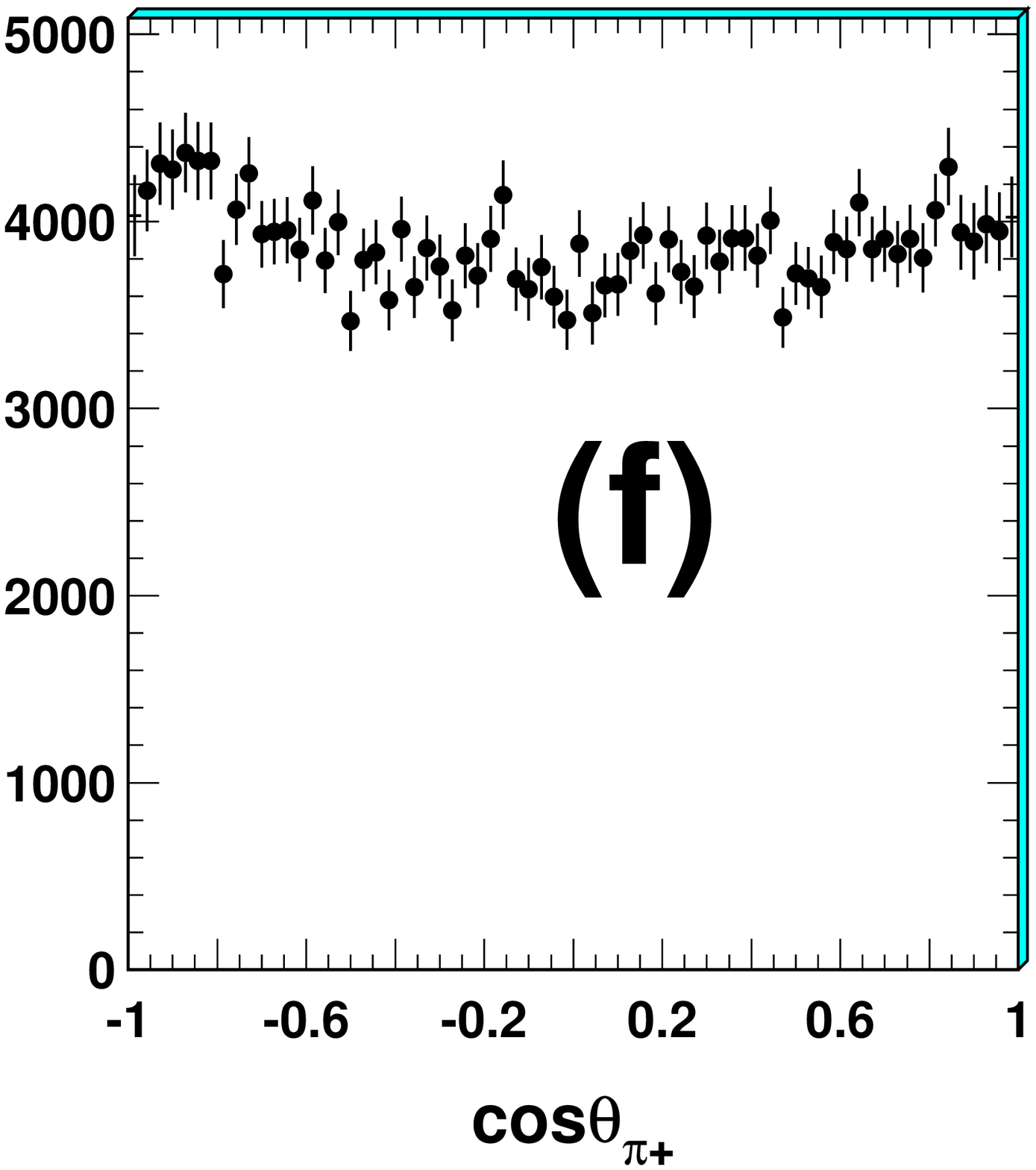, width=4.5cm,height=4.0cm}}
\caption{Fit results of
$\psi(2S)\rightarrow\pi^+\pi^-J/\psi$~(P.K.U. ansatz). Dots with
error bars are data and the histograms are the fit results.
 (a) and (b) are the  $\pi^+\pi^-$ invariant mass,
 (c) and (d) the cosine of the $\sigma$ polar angle in the lab frame,
 and (e) and (f) the cosine of the $\pi^+$ polar angle in the $\sigma$
 rest frame.  The upper plots  are the detected distributions, while the
 bottom ones are the distributions after efficiency correction.
}
\label{mass-just}
\end{figure*}


\begin{figure*}[htbp]

 \centerline{
\psfig{file=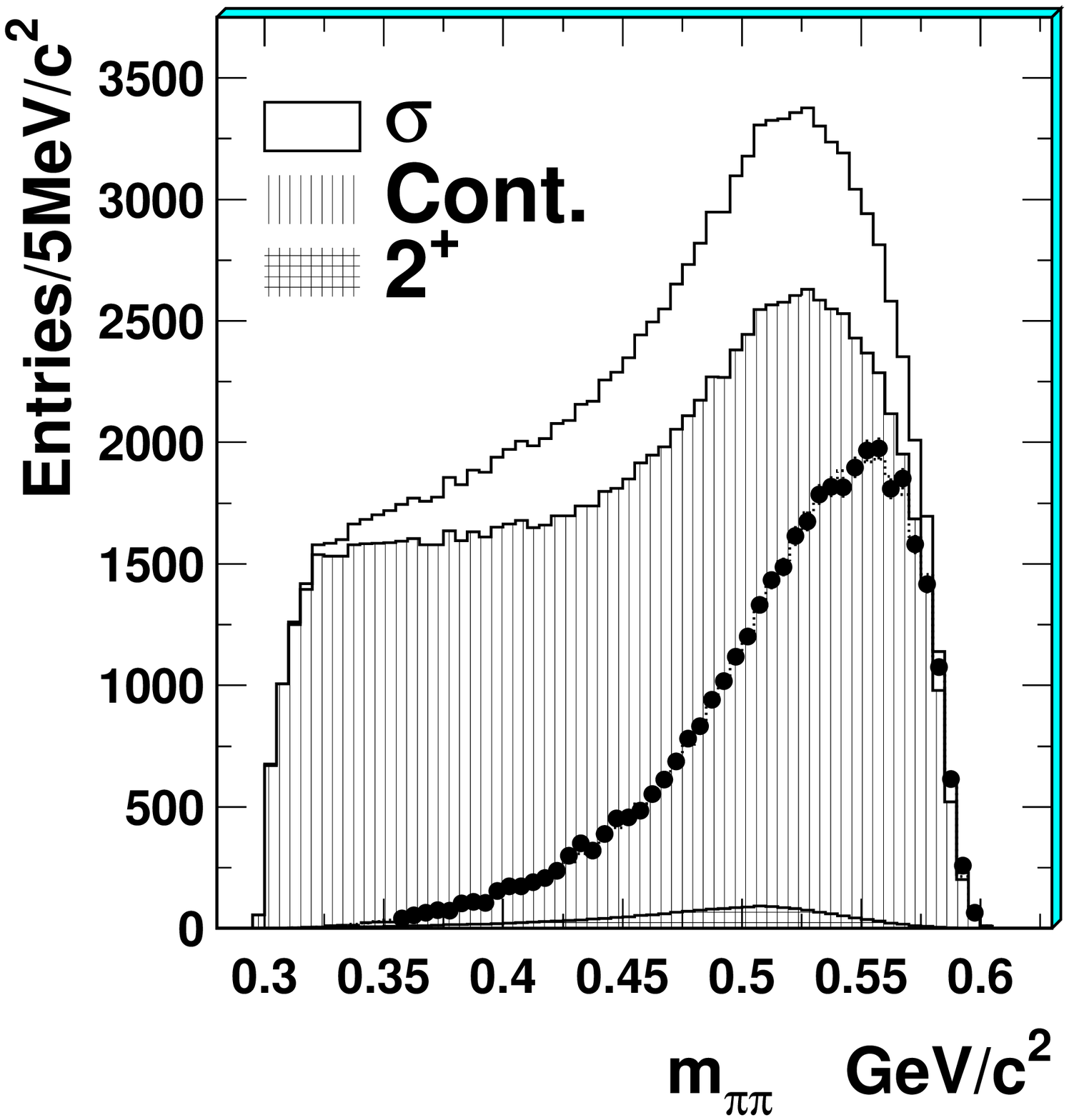, width=6cm,height=5.0cm}
\psfig{file=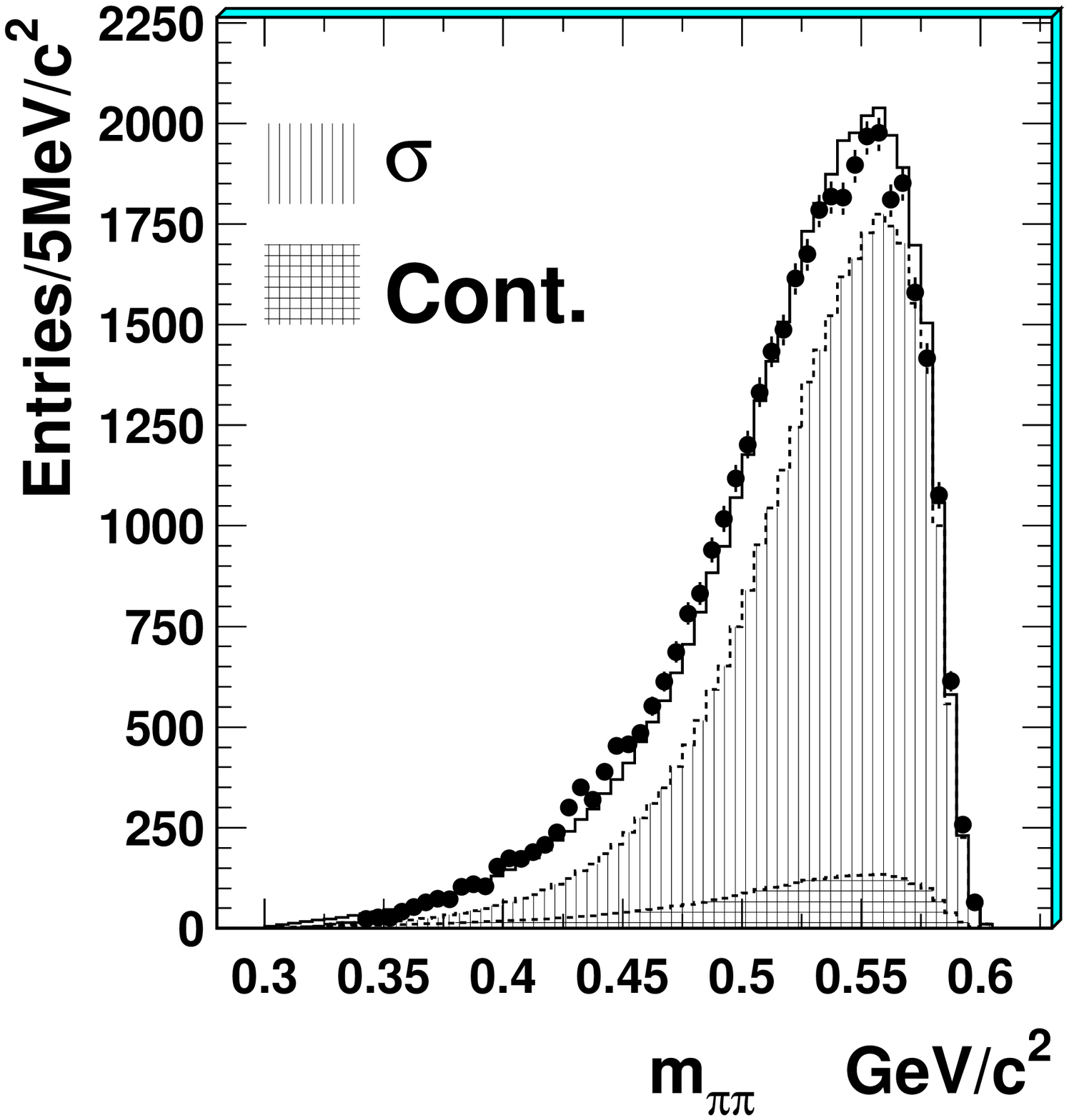,width=6cm,height=5.0cm}
}
\caption{ The $\pi^+\pi^-$ invariant mass distribution including the
  components. The left is the P.K.U. ansatz, which has the
  contributions from $\sigma$, $D$-wave term, contact term, and
  their sum (the $D$-wave is enlarged by a factor of 20 in the
  figure).  The right is the $\pi^+\pi^-$ invariant mass fitted by the
  formula from Ref.{~\cite{guofk,oset}}, with no explicit $D$-wave.
  Dots with error bars are data, and the histograms are the fit
  results.}
\label{mass-guofk}
\end{figure*}


The global fits determine the best estimation
of the Breit-Wigner parameters
for each parameterization.
The pole position in the complex energy plane is
related to the mass and width of the resonance by
\begin{eqnarray}
\sqrt{s_{pole}}=m_{\sigma}-i\frac{\Gamma_{\sigma}}{2}~.
\label{pole-1}
\end{eqnarray}

The best fit results and the corresponding pole positions for all the
parameterizations are listed in Table {\ref{scan-1}}. The statistical
error of the resonance mass~(width) is determined by a decrease of
$\frac{1}{2}$ in the log-likelihood from its maximum value with all
other parameters fixed to their best solutions.

For the second model, where the VPP vertex is represented by an
effective Lagrangian and the $\pi\pi$ $S$-wave FSI is included, the
$\pi\pi$ mass spectrum can also be reproduced well.
Here the $\sigma$ requires a much
smaller interference between the S-wave FSI and
the contact term.
In this case, the fit is worse than the fits of the first model;
this may due to the fixed pole position of the $\sigma$ and the neglected
$D$-wave contribution. For the second model,
the pole is not measured in the fit, but taken from
Ref.{~\cite{oset}}, which was determined from $\pi\pi$ scattering
data.

\begin{table*}[hptb]
\caption{ Fit results of the two models and all
the Breit-Wigner parameterizations}
 \label{scan-1} \centerline{\footnotesize
\begin{tabular}{c|ccccc} \hline
Model    &  Constant $\Gamma$ &  $\Gamma$ with $\rho$ &
P.K.U.ansatz &Bugg \& Zou's approach & Ref. {~\cite{guofk,oset}}\\\hline
&~~~$(553\pm15 \pm 47)$&~~~$(559\pm~6 \pm 26)$
&~~~$(554\pm13\pm 65)$ &~~~$(541\pm9 \pm 95)$ & {$469-i203$}\\
\raisebox{5pt}{Pole(MeV/$c^2$)}& $-i(254\pm23 \pm 54)$& $-i(179\pm~7 \pm
18)$ & $-i(240\pm~4\pm 19)$ &$-i(253\pm8 \pm 33)$ &  (input)
\\
$N_\sigma{\dagger}$           &140308&72735&133208&171586&30765 \\
$N_{\mbox{contact}}{\dagger}$ &121625&63133&111230&157741&3039 \\
-ln{\cal L}     &-16174.5         &-16166.8
&-16171.0   &-16174.3   &-15974.1  \\
$\chi^2_{obs.}/ndf$&217.83/196    &227.54/196        &224.07/196 &
217.88/196&392.73/208\\
C.L.            &0.1362
&0.0608            &0.0825     & 0.1357 &$3\times 10^{-13}$
\\\hline
\end{tabular}}
$\dagger$ : $N_\sigma$ and $N_{contact}$ are the numbers of events
in the fit.
\end{table*}

To check the goodness of fit in our analysis, we construct a
variable
\begin{eqnarray}
\chi^2_{obs}=\sum_{i=1}^{N} \left(\frac{N_{i}^{DT}-N_{i}^{MC}}
{\sqrt{N_{i}^{DT}}}\right)^2~, \label{chi2-1}
\end{eqnarray}
where $N$ is the number of cells, $N_{i}^{DT}$ and $N_{i}^{MC}$ are
the numbers of events in the $i$'th cell of the Dalitz plot with
axes $m^{2}_{\pi^+\pi^-}$ and $m_{J/\psi \pi^+}^{2}$ for data
and MC simulation, respectively. Such a variable should be distributed
according to the
$\chi^2$ distribution with $n=N-K$ degrees of freedom, where $K=12$ is
the number of parameters to be determined in our Maximum Likelihood
fit. In our case, 15 bins in both $m^{2}_{\pi^+\pi^-}$ and $m_{J/\psi
\pi^+}^{2}$ give 225 cells. To ensure proper $\chi^2(n)$ behavior,
cells with less than five events have been merged into adjacent ones.
The number of cells becomes $N=208$, and the number of degrees of
freedom $n=196$.  From the observed $\chi^2$ value determined using
Eq.~(\ref{chi2-1}) for each parameterization, the confidence
levels~(C.L.) are calculated and listed in Table {\ref{scan-1}}.

\section{\boldmath Systematic Errors}
For the first model,
the systematic error of the $\sigma$ pole position
arises from the uncertainties of the strength of
the $2^+$ component, the form of the contact term,
and the inconsistency between data and of MC simulation.
For the $2^+$ component,
we conservatively remove it from the fit, and the
difference of the fitted values from the nominal values are taken
as systematic errors. Two contact terms,
namely, constant amplitude and $\alpha_1+i\alpha_2\rho$, where
$\alpha_{1}$  and $\alpha_{2}$ are two parameters to be fitted,
are adopted in the fit; the difference is considered as the
systematic error. The MC simulation and data have
different mass resolutions in the high mass region of
the $\pi^+\pi^-$ system.
A modification of $\pi^+\pi^-$ mass resolution is made to improve the fit,
and the difference of the fitted pole positions with and without
this modification is taken as the systematic error. The
systematic error from non-signal backgrounds is neglected.
\begin{table*}[htbp]
\caption{Systematic errors
in the pole position (~\mbox{MeV}/$c^2$)}
\vspace{0.0cm}
\label{error-ana-1}
\centerline{\small
\begin{tabular}{c|cc|cc|cc|cc} \hline
&\multicolumn{2}{|c}{Constant $\Gamma$}
&\multicolumn{2}{|c}{$\Gamma$ with $\rho$}
&\multicolumn{2}{|c}{P.K.U. ansatz}
&\multicolumn{2}{|c}{Bugg \& Zou's approach}  \\
&$\sigma_{m}$&    $\sigma_{\Gamma/2}$&
 $\sigma_{m}$&    $\sigma_{\Gamma/2}$&
 ~~~$\sigma_{m}$~~~&    $\sigma_{\Gamma/2}$&
 ~~~~$\sigma_{m}$~~~~&    $\sigma_{\Gamma/2}$\\
\hline
 $2^{+}$ uncertainty & 22 &35 &  4 & 14& 10 & 2 &  3 & 3   \\
 form of contact term& 11 &14 &  1 &  2& 30 & 8 &  91& 22  \\
 M.C. imperfection& 40 &38 & 25 & 10& 56 &17 &  37& 24  \\
 Total Error         & 47 &54 & 26 & 18& 65 &19 &  95& 33  \\\hline
\end{tabular}}
\end{table*}

In order to obtain the $m_{\sigma}$ and
$\frac{\Gamma_{\sigma}}{2}$ errors in Eq. (\ref{pole-1}), we set the
denominator in Eq.~(\ref{bw0}) equal to zero and obtain the
pole position and corresponding errors by taking into account
the  mass and width errors of the Breit-Wigner
parameterizations. This is done using a MC
sampling method, where the correlation between the mass and width
is ignored. Table {\ref{error-ana-1}} summarizes
the systematic errors  from all sources,
and Table {\ref{scan-1}} lists the parameters of pole position
 and their total errors.


\section{\boldmath Results and discussion}

The process $\psi(2S)\rightarrow \pi^+\pi^-J/\psi, ~J/\psi\rightarrow
\mu^+\mu^-$ is studied based on $14\times 10^6$ $\psi(2S)$ events
collected with the BESII detector.  The $\pi^+\pi^-$ invariant mass
spectrum of $\psi(2S)\rightarrow \pi^+\pi^-J/\psi$ has a severe
suppression near the $\pi^+\pi^-$ threshold, which is distinctly
different from phase space and suggests $\sigma$ production in the
process.  We fit the data with two different models. For the first
model, using four different Breit-Wigner parameterizations, the data
can be well fitted, although a strong cancellation between the
$\sigma$ and the contact term is required. In fact, such a large
cancellation is dictated by chiral symmetry~\cite{cahn,ishida}. The
pole positions of $\sigma$ are determined for different Breit-Wigner
parameterizations, which are
$(553\pm15\pm47)-i(254\pm23\pm54)$~MeV$/c^2$(constant width), $(559\pm
6 \pm 26)-i(179 \pm 7 \pm 18)$~MeV$/c^2$~(width containing kinematic
factor), $(554\pm 13 \pm 65)-i(240\pm 4\pm
19)$~MeV$/c^2$~(P.K.U. ansatz), and $(541\pm 9 \pm 95)-i(240\pm 8\pm
33)$~MeV$/c^2$~(Bugg \& Zou's approach). The first
Breit-Wigner parameterization may be problematic because the imaginary
part does not vanish at threshold.  The
second parameterization gives a small $\sigma$ width, and creates a
virtual state in the real energy axis below the $\pi\pi$
threshold{~\cite{zheng-hq-1}}. The final best estimate of the $\sigma$
pole position from this analysis is
$(552^{+~84}_{-106})-i(232^{+81}_{-72})$~MeV$/c^2$, where the central
values are obtained by a simple mean of the different
Breit-Wigner parameterizations excluding the second one, while the errors
cover the statistical and systematic errors, including the
differences in the three Breit-Wigner parameterizations.

We also fit our data according to the scheme in
Ref.{~\cite{guofk}}. It is found that {the $\pi\pi$ S-wave FSI plays a
dominant role in $\psi(2S)\rightarrow\pi^+\pi^-J/\psi$, while the
contribution from the contact term is small.  This means that the
$\sigma$ meson has a significant contribution in this process. The
$\sigma$ pole used in this fit, $469-i203$~MeV/$c^2$ is consistent
with the fits to the Breit-Wigner functions.  This implies that,
although the two theoretical schemes are very different, both of them
find the $\sigma$ meson at similar pole positions.

If the $\sigma$ meson exists, the pole should occur universally in all
$\pi\pi$ system with correct quantum numbers. Our analysis
demonstrates that, in $\psi(2S)\rightarrow\pi^+\pi^-J/\psi$, even
though there is no apparent peak structure, one can still determine
the pole location in good agreement with that obtained from
$J/\psi\rightarrow \omega\pi^+\pi^-$ decay {~\cite{ref_41}} by
assuming a simple form of the contact term. Hence it provides further
evidence for the $\sigma$ meson.




\section*{\boldmath Acknowledgment}
The BES collaboration thanks the staff of BEPC and
computing center for their hard efforts. We are grateful to F.K. Guo
and P.N. Shen for useful discussions and good suggestions.
This work is supported in
part by the National Natural Science Foundation of China under
contracts Nos. 10491300, 10225524, 10225525, 10425523, the Chinese
Academy of Sciences under contract No. KJ 95T-03, the 100 Talents
Program of CAS under Contract Nos. U-11, U-24, U-25, and the
Knowledge Innovation Project of CAS under Contract Nos. U-602,
U-34 (IHEP), the National Natural Science Foundation of China
under Contract No. 10225522 (Tsinghua University), and the
Department of Energy under Contract No.DE-FG02-04ER41291 (U
Hawaii).



\end{document}